\documentclass[twocolumn,runningheads]{svjour2}
\smartqed  
\usepackage{graphicx}
\journalname{Astrophysics and Space Science}
\begin{document}
\def\la{\mathrel{\hbox{\rlap{\hbox{\lower4pt\hbox{$\sim$}}}\hbox{$<$}}}}
\def\ga{\mathrel{\hbox{\rlap{\hbox{\lower4pt\hbox{$\sim$}}}\hbox{$>$}}}}

\title{The Blazar Sequence: Validity and Predictions
}


\author{Paolo Padovani}


\institute{P. Padovani \at
              European Southern Observatory, Karl-Schwarzschild-Str. 2\\
              D-85748 Garching bei M\"unchen, Germany \\
              Tel.: +49-89-32006478\\
              Fax: +49-89-32006677\\
              \email{ppadovan@eso.org}           
    }

\date{Invited talk at the Workshop "The Multi-messenger approach to high energy gamma-ray sources",
Barcelona, July 2006}


\maketitle

\begin{abstract}
The "blazar sequence" posits that the most powerful BL Lacertae objects and
flat-spectrum radio quasars should have relatively small synchrotron peak
frequencies, $\nu_{\rm peak}$, and that the least powerful such objects
should have the highest $\nu_{\rm peak}$ values. This would have strong
implications for our understanding of jet formation and physics and the
possible detection of powerful, moderately high-redshift TeV blazars. I
review the validity of the blazar sequence by using the results of very 
recent surveys and compare
its detailed predictions against observational data. I find that the blazar 
sequence in its simplest form is ruled out. However, powerful flat-spectrum radio
quasars appear not to reach the $\nu_{\rm peak}$ typical of BL Lacs. This
could indeed be related to some sort of sequence, although it cannot be
excluded that it is instead due to a selection effect.

\end{abstract}

\section{Introduction}
\label{intro}

Blazars are the most extreme variety of Active Galactic Nuclei (AGN)
known. Their signal properties include irregular, rapid variability; high
optical polarization; core-dominant radio morphology; apparent superluminal
motion; flat ($\alpha_{\rm r} \la 0.5$; $f_\nu \propto \nu^{-\alpha}$)
radio spectra; and a broad continuum extending from the radio through the
gamma-rays \cite{UP95}. Blazar properties are consistent with
relativistic beaming, that is bulk relativistic motion of the emitting
plasma at small angles to the line of sight, which gives rise to strong
amplification and collimation in the observer's frame. The blazar class
includes flat-spectrum radio quasars (FSRQ) and BL~Lacertae objects, which
are thought to be the ``beamed'' counterparts of high- and low-luminosity
radio galaxies, respectively. That is, according to unified schemes blazars
are simply radio galaxies with their radio jets forming a small angle with
respect to the line of sight \cite{UP95}.  This also explains the intrinsic rarity
of the blazar class.

Two blazar properties are most interesting for this paper and this
conference: 1. their spectral energy distributions (SEDs), which are
usually explained in terms of synchrotron and inverse Compton emission, the
former dominating at lower energies, the latter being relevant at higher
energies; 2. the fact that blazars are sites of very high energy phenomena,
with bulk Lorentz factors up to $\sim 40$ (corresponding to velocities 
$\sim 0.9997$c) and photon energies reaching the TeV range. As a 
consequence, and despite their small numbers, blazars dominate the high energy
sky. Indeed, the large majority of extragalactic sources detected by EGRET
are blazars (see various papers at this conference), while 12/13 of the
extragalactic TeV sources detected so far are BL Lacs \cite{maz07}.

The main difference between the two blazar classes lies in their emission
lines, which are strong and quasar-like for FSRQ and weak or in some cases
outright absent in BL~Lacs. Another difference between the two classes,
which has been a puzzle for quite some time, relates to their SED. BL Lacs
have a large range in synchrotron peak frequency, $\nu_{\rm peak}$, which
is the frequency at which the synchrotron energy output is maximum (i.e.,
the frequency of the peak in a $\nu - \nu f_{\nu}$ plot). Although the
$\nu_{\rm peak}$ distribution appears now to be continuous, it is still
useful to divide BL Lacs into low-energy peaked (LBL), with $\nu_{\rm
peak}$ in the IR/optical bands, and high-energy peaked (HBL) sources, with
$\nu_{\rm peak}$ in the UV/X-ray bands \cite{pg95}. The location of the
synchrotron peaks suggests in fact a different origin for the X-ray
emission of the two classes. Namely, an extension of the synchrotron
emission responsible for the lower energy continuum in HBL, which display
steep ($\alpha_{\rm x} \sim 1.5$) X-ray spectra, and inverse Compton
emission in LBL, which have harder ($\alpha_{\rm x} \sim 1$) spectra
\cite{pad01,pad04,wol98}. Given their high-energy peak, HBL are more easily
selected in the X-ray band and have relatively large X-ray-to-radio flux
ratios, $f_{\rm x}/f_{\rm r}$. LBL, on the other hand, have smaller $f_{\rm
x}/f_{\rm r}$ values and are typically selected in the radio band.

The puzzling thing was that no such distinction appeared for FSRQ. All
known FSRQ were of the ``L'' type, i.e., with low (IR/optical energies)
$\nu_{\rm peak}$ and, therefore, X-ray band dominated by inverse Compton
emission.  No ``HFSRQ'' (as these sources have been labelled), i.e., FSRQ
with high (UV/X-ray energies) $\nu_{\rm peak}$ seemed to exist until a few
years ago.

It is important to point out up front that the determination of $\nu_{\rm peak}$ is not
straightforward, as it requires building the SED with a decently large
number of data points across a large energy range. Moreover, being blazars
variable at all wavelengths, the non-simultaneous SEDs which are quite
often used imply a further uncertainty on $\nu_{\rm peak}$. A "proxy" which
is sometimes adopted is $f_{\rm x}/f_{\rm r}$, which, as mentioned above,
depends on $\nu_{\rm peak}$ (equivalently, the effective radio--X-ray
spectral index $\alpha_{\rm rx}$, normally defined between 5 GHz and 1 keV,
is also used). For consistency with previous work, which we refer to
extensively, we have adopted the values $H_0 = 50$ km/s/Mpc and $q_0 =0$.

\section{The Blazar Sequence}
\label{sequence}

The so-called "Blazar Sequence" was proposed in 1998 by two papers
\cite{fos98,ghi98}. One of the main results of \cite{fos98} is given in
their Fig. 7, which plots various powers vs. $\nu_{\rm peak}$ for three
blazar samples: the 2 Jy FSRQ (radio-selected), the 1 Jy BL Lacs
(radio-selected), and the {\it Einstein} Slew Survey BL Lacs (X-ray
selected). An anti-correlation was apparent, with the most powerful sources
having relatively small synchrotron peak frequencies and the least powerful
ones having the highest $\nu_{\rm peak}$ values.

The theoretical interpretation to this anti-correlation was given by
\cite{ghi98}. The frequency of the peak of the synchrotron emission is
related to the electron energy, as $\nu_{\rm peak} \propto B \delta
\gamma_{\rm peak}^2$, where $B$ is the magnetic field, $\delta$ is the
Doppler factor, and $\gamma_{\rm peak}$ is a characteristic electron energy
which is determined by a competition between accelerating and cooling
processes. Since in more powerful sources the energy density ($U \propto
L/R^2$, where $R$ is the characteristic size of the jet) is higher, the
emitting particles have a larger probability of losing energy and therefore
are subjected to more cooling. This translates into a lower value of
$\gamma_{\rm peak}$ and therefore of $\nu_{\rm peak}$. Fig. 7 of
\cite{ghi98} summarizes the blazar sequence by plotting $\gamma_{\rm peak}$
vs. energy density, $U$, for HBL (high $\gamma_{\rm peak}$ - low $U$), LBL
(lower $\gamma_{\rm peak}$ - higher $U$), and FSRQ (HPQ and LPQ in their
notation; low $\gamma_{\rm peak}$ - high $U$).

\subsection{Predictions and Physical Implications}
\label{predictions}

The blazar sequence makes very specific predictions. Na\-me\-ly:

\begin{enumerate}

\item since FSRQ are more powerful than BL Lacs (which is explained by the
fact that FSRQ are thought to be the beamed version of high power
radio-galaxies, the so-called Fanaroff-Riley type II), an anti-correlation
between power and $\nu_{\rm peak}$ implies that FSRQ with high $\nu_{\rm
peak}$ should not exist. This would then explain the puzzle of the missing
HFSRQ. The immediate implication is that, since all known BL Lacs with TeV
detections (twelve as of this meeting) are of the HBL type, TeV detectors
should not expect to observe any FSRQ. A complication is that TeV photons
interact with background infrared photons to produce electron-positron
pairs and get therefore depleted, the more so the larger the distance of
the emitter. Nevertheless these sources, if they existed, would increase
the statistics and, being at higher redshifts than BL Lacs, would better
constrain the IR background and, therefore, the star formation history in
the Universe \cite{cos07};

\item since low-luminosity sources are more numerous than high-luminosity
ones (as all observed luminosity functions are of the type $\phi(L) \propto
L^{-\alpha}$, with $\alpha > 0$), an anti-correlation between power and
$\nu_{\rm peak}$ implies that HBL should be more numerous than LBL. The
physical implications are two-fold: a) a simple demographical one relevant,
for example, for deep surveys; b) strong constraints on jet physics. In
fact, as $\nu_{\rm peak} \propto B \delta \gamma_{\rm peak}^2$, if high
$\nu_{\rm peak}$ values were indeed more common, this would mean that
Nature prefers certain types of jets and therefore some special combination
of these parameters, a fact certainly worth of a thorough investigation.

\end{enumerate}

\subsection{Tests}
\label{tests}

These very specific predictions lend themselves to be tested, at least in
theory, relatively simply. Namely, one can prove or disprove the sequence
by:

\begin{enumerate}

\item checking the power - $\nu_{\rm peak}$ anti-correlation;

\item finding any "forbidden" objects, that is outliers from the correlation 
(high $\nu_{\rm peak}$ - high power and/or low $\nu_{\rm peak}$ - low
power blazars); 

\item counting sources; that is, are HBL really more numerous than LBL?  
(and is this consistent with the X-ray background?)

\end{enumerate}

I will discuss these tests in detail in the following. 

\begin{figure*}
\centering
  \includegraphics[width=0.75\textwidth]{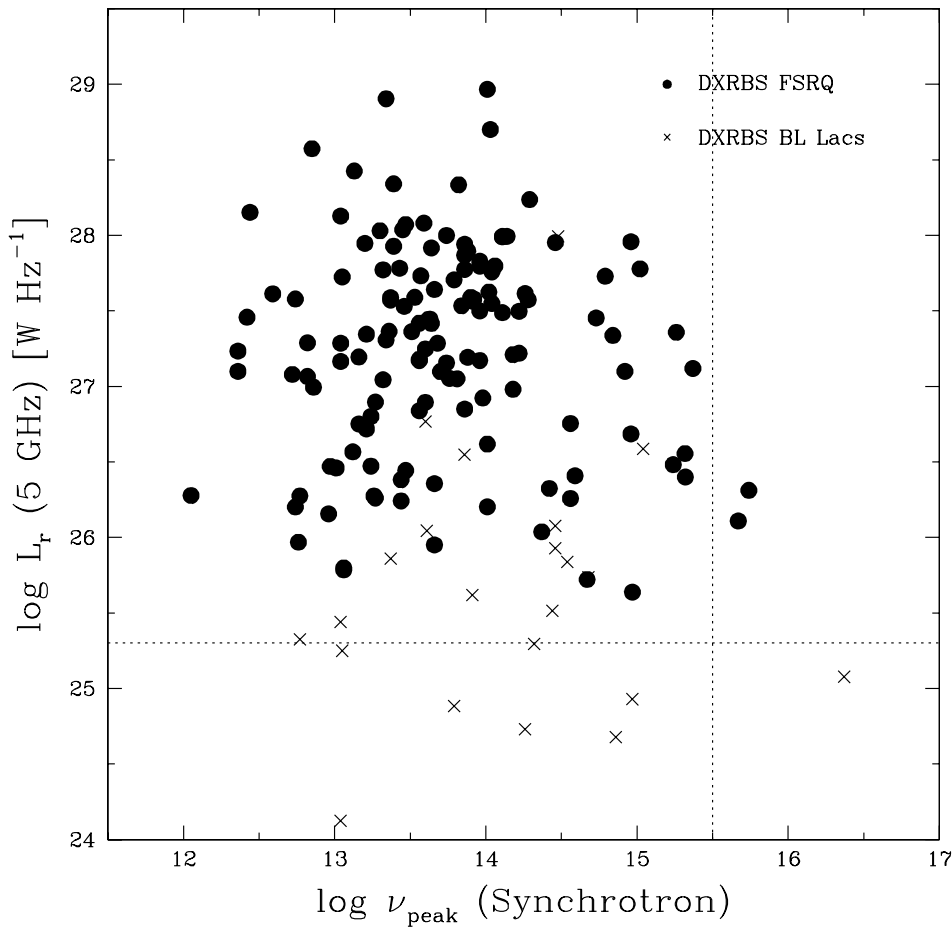}
\caption{Radio power at 5 GHz vs. the synchrotron peak frequency for FSRQ
(filled points) and BL Lacs (crosses) for the DXRBS sample. The dotted lines
denote the two quadrants (top-left and bottom-right) occupied by the sources
studied by \cite{fos98}.}
\label{lrnupeak}  
\end{figure*}

\section{The power - $\nu_{\rm peak}$ anti-correlation}
\label{correlation}

Before discussing any test of the anti-correlation, it is important to see
in detail how the original plot was derived. Two BL Lac samples had been
used by \cite{fos98}, one radio- and one X-ray-selected, and one FSRQ
sample. These samples had been assembled in an independent and somewhat
different way, especially so as regards the selection band. Two caveats
need then to be kept in mind: 1. it is always dangerous to infer parameter
dependencies by plotting samples selected in an inhomogeneous way,
particularly if one of the parameters depends on the selection method as in
this case (most HBL are X-ray selected while most LBL are
radio-selected). Indeed, none of the individual samples shown in Fig. 7 of
\cite{fos98} showed the claimed anti-correlation between power and
$\nu_{\rm peak}$, which was only apparent by combining the three samples;
2. the only FSRQ sample was radio-selected.  As the objects with the
largest $\nu_{\rm peak}$ in the plot were X-ray selected BL Lacs, one might
argue that the lack of high $\nu_{\rm peak}$ - high power sources was due
to the lack of X-ray selected FSRQ.

In any case, it is clear that an independent check for the existence of
this anti-correlation needed to be carried out. This has been done by
various groups, whose results I am going to review next in chronological
order.

The Deep X-ray Radio Blazar Survey (DXRBS) uses a double X-ray/radio
selection and contains mostly FSRQ \cite{pad03,pad06}. DXRBS is at present
the faintest and largest flat-spectrum radio sample with nearly complete
($\sim 95\%$) identifications down to fluxes $10-20$ times fainter than
previous radio and X-ray surveys \cite{per98,lan01}. Therefore, it obviates
to the selection effects present in the samples used by \cite{fos98}. The
DXRBS radio power - $\nu_{\rm peak}$ plot is shown in Fig. \ref{lrnupeak},
which shows no correlation between the two parameters, a huge scatter,
reaching 4 orders of magnitude in power, and outliers, that is sources
occupying regions of this plot which were empty in the original one by
\cite{fos98}. In particular, of the 21 BL Lacs with $\nu_{\rm peak} <
10^{15.5}$ Hz and redshift information, $\sim 1/3$ "invade" the low-power
part ($L_{\rm r} < 10^{25.3}$ W/Hz) of the plot.

The CLASS blazar survey has been used by \cite{cac04} to study the radio
power - $\alpha_{\rm rx}$ correlation. As mentioned in Sect. \ref{intro},
this latter parameter is a proxy for $\nu_{\rm peak}$. Their Fig. 7 shows
that, contrary to the predictions of the blazar sequence, many sources at
relatively low power and with $\alpha_{\rm rx} > 0.75$ (that is, relatively
large $f_{\rm x}/f_{\rm r}$) were found. In other words, even the CLASS
sample shows the presence of low-power--low-$\nu_{\rm peak}$ BL Lacs. One
complication with this result, however, is the fact that the relationship
between $\alpha_{\rm rx}$ and $\nu_{\rm peak}$ is not very tight (see,
e.g., Fig. 11 of \cite{pad03}).

The 200 mJy sample was used by \cite{ant05} to study the radio power - 
$\nu_{\rm peak}$ correlation. Their Fig. 4 shows a large number of sources with
$\nu_{\rm peak} < 10^{15.5}$ Hz and $\nu L_{\rm 5GHz} < 10^{42}$ erg/s,
whereas none were found in \cite{fos98}. Moreover, there is no correlation
between the two parameters and a large scatter, reaching almost 5 orders of
magnitude in power at a given $\nu_{\rm peak}$, is present.

\begin{figure*}
\centering
  \includegraphics[width=0.75\textwidth]{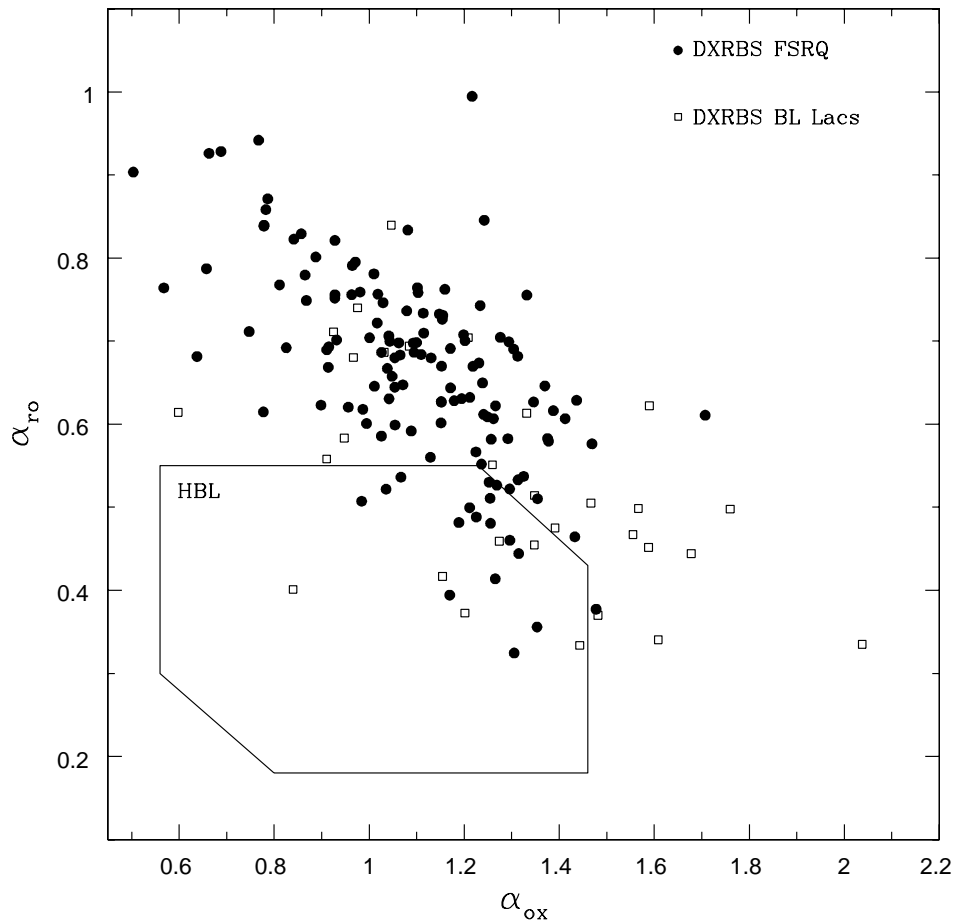}
\caption{($\alpha_{\rm ro}$, $\alpha_{\rm ox}$) plane for the DXRBS
sample. Effective spectral indices are defined in the usual way and
calculated between the rest-frame frequencies of 5 GHz, 5,000 \AA, and 1
keV . Filled circles represent FSRQ, while open squares represent BL
Lacs. The region in the plane within $2\sigma$ from the mean $\alpha_{\rm
ro}$, $\alpha_{\rm ox}$, and $\alpha_{\rm rx}$ values of HBL is indicated by a
solid polygon.}
\label{aroaox}  
\end{figure*}

The SEDs for a large, heterogeneous sample of BL Lacs taken from the
Veron-Cetty \& Veron BL Lac catalogue \cite{ver00} and visible from the
Metsh\"ahovi radio observatory were assembled by \cite{nie06}. Their Fig. 3
shows an anti-correlation between radio power and $\nu_{\rm peak}$, a huge
scatter, reaching 5 orders of magnitude in power, and many outliers as
compared to Fig. 7 of \cite{fos98}, especially in the
low-power--low--$\nu_{\rm peak}$ region.

The situation of the radio power - $\nu_{\rm peak}$ anti-cor\-re\-la\-tion, as
inferred by these studies, can then be thus summarised:

\begin{enumerate}

\item No radio power - $\nu_{\rm peak}$ anti-correlation is present when
homogeneous, well-defined samples are used. Only when putting together
objects from various surveys is such an anti-correlation observed. This
points to selection effects being at the origin of the original
anti-correlation;

\item For all studies, the scatter is huge, reaching 4-5 orders of
magnitude in power at a given $\nu_{\rm peak}$. Therefore, even if there 
were an anti-correlation, it could not be very tight;

\item Outliers, that is sources which occupy regions of the plot which were
empty in the original version, have been found by all studies. It is fair
to say, though, that these are mostly in the low-power--low--$\nu_{\rm
peak}$ region.

\end{enumerate}

This last point is quite relevant. When using samples at lower fluxes than
those used by \cite{fos98}, one is sampling sources which, by being
fainter, could also be less beamed and therefore less powerful. Hence, one
could expect some of the sources above the horizontal dotted line in
Fig. 7 of \cite{fos98} (see also Fig. \ref{lrnupeak}) to move below the line 
in deeper surveys \cite{ghi03}.
It is then especially important to look for the other type of outliers, the
HFSRQ, that is high-power--high--$\nu_{\rm peak}$ blazars.

\section{Looking for HFSRQ}
\label{hfsrq}

How does one look for HFSRQ, that is high-power blazars with high $\nu_{\rm
peak}$?  The steps are, at least in theory, simple enough. Namely:

\begin{enumerate}

\item start from X-ray selected samples, as HBL are mostly found in the
X-ray band;

\item select suitable candidates by looking in regions of parameter space
which are known to be occupied by HBL. For example, those defined by large
$f_{\rm x}/f_{\rm r}$;

\item after pre-selection, build the SED. One should check that, since we
are dealing with quasars, which are known for their ultraviolet excess, any
high-energy synchrotron peak is not due to the ultraviolet "bump";

\item as a final step, confirmation by X-ray observations is
recommended. As discussed in Sect. \ref{intro}, in fact, for a source to be
an HFSRQ its X-ray spectrum should be synchrotron dominated and, therefore,
relatively steep ($\alpha_{\rm x} > 1$) or, at least, concave (which would
suggest that the X-ray band is sampling the synchrotron to inverse Compton
transition).

\end{enumerate}

However, all of the above is quite complex and time-consuming to put into
practice. In fact, to the best of my knowledge, only our group has tackled
this problem to this level of details. This is described partly in
\cite{pad02,pad03}, to which the reader is referred for more details. In
brief, as an initial step towards studying the broad-band properties of our
sources, we first derived their $\alpha_{\rm ox}$, $\alpha_{\rm ro}$, and
$\alpha_{\rm rx}$ values. These are the usual rest-frame spectral indices
defined between 5 GHz, 5,000 \AA, and 1 keV. The fraction of sources which
fall in the region of the plane within $2\sigma$ from the mean $\alpha_{\rm
ro}$, $\alpha_{\rm ox}$, and $\alpha_{\rm rx}$ values of HBL, the ``HBL
box'', derived by using all HBL in the multi-frequency AGN catalog of
\cite{pad97}, is $\sim 15\%$ and $\sim 9\%$ for DXRBS BL Lac objects and
FSRQ respectively (see Fig. \ref{aroaox}). This already shows that $\sim
10\%$ of DXRBS FSRQ have broad-band colours typical of high-energy peaked BL
Lacs. For comparison, the 1 Jy FSRQ, a {\it radio-selected} sample, occupy
a region of $\alpha_{\rm ox},\alpha_{\rm ro}$ parameter space with
$\alpha_{\rm rx}$ similar to that typical of LBL. FSRQ with low
$\alpha_{\rm rx}$ ($\la 0.78$, roughly equivalent to the HBL/LBL division) 
constitute only $\sim 5\%$ of the 1 Jy sources with X-ray data. However, 
none of the 1 Jy FSRQ fall in the HBL box.

{\it BeppoSAX} observations of four candidate HFSRQ were carried out by
\cite{pad02} with mixed results: one source had an X-ray spectrum dominated
by inverse Compton emission, while two others had a flat X-ray spectrum
with evidence of steepening at low energies. RGBJ1629\-+4008, however, was
clearly synchrotron dominated in the X-ray band, with $\alpha_{\rm x} \sim
1.5$, which is typical of HBL. This source represents therefore the first
example of confirmed HFSRQ, with $\nu_{\rm peak} \sim 2 \times 10^{16}$ Hz
($\sim 0.1$ keV).

It has to be pointed out, however, that, despite being an FSRQ, its
relatively low radio power ($L_{\rm 5GHz} \sim 6 \times 10^{24}$ W/Hz)
places this source still in the bottom-right quadrant of Fig. \ref{lrnupeak}, that is, where
it should be according to the blazar sequence. Moreover, although clearly way 
above the values reached by LBL and well
within the HBL range, its $\nu_{\rm peak}$ is towards the low end of the
HBL distribution.

The latter result has been confirmed by more extensive searches, summarized
in \cite{pad03,gio05,lan06b}. In short, various X-ray selected samples have
been searched thoroughly, and XMM and Chandra data have been taken
\cite{lan06b}.  HFSRQ, that is, broad-lined blazars with $\nu_{\rm peak}$
typical of HBL have been found, despite previous claims to the contrary
(the issue of the effect of the UV bump on the estimated value of $\nu_{\rm
peak}$ has been discussed by \cite{pad03,lan06b}). However, their maximum
$\nu_{\rm peak}$ is still $\sim 0.1$ keV, while HBL reach {\it typically}
$\sim 1$ keV and {\it exceptionally} $\sim 100$ keV (as in the case of MKN
501; \cite{pia98})

The most extreme case is that of the Sedentary survey \cite{gio99,gio05},
which was designed to find extreme HBL, as a selection was made on $f_{\rm
x}/f_{\rm r}$ (roughly equivalent to $\nu_{\rm peak} \ga 10^{16}$
Hz). Nineteen broad-lined AGN were found, out of 169 candidates, as
compared to 150 HBL, but none of them was a definite FSRQ. These sources,
in fact, turned out to be mostly nearby, low radio luminosity AGN very
close to the radio-loud/radio-quiet border.

Is this lack of HFSRQ with high $\nu_{\rm peak}$ values telling us
something about jet physics and the blazar sequence or is it still a
selection effect? Recall that RGBJ1629+4008, and also the other three HFSRQ
candidates studied in \cite{pad02}, had all relatively low radio powers
($\langle L_{\rm 5GHz} \rangle \sim 10^{25}$ W/Hz), more typical of BL Lacs
than of FSRQ and in any case close to the low-luminosity end of the FSRQ
radio luminosity function \cite{pad06}. Similarly, the HFSRQ studied by
\cite{lan06a} were also of low power, reaching powers more typical of
low-luminosity (Fanaroff-Riley type I) radio ga\-la\-xies.

One could argue that the fact that HFSRQ are found at relatively low radio
power is not coincidental. Indeed, for a high radio power HBL-like source, 
the optical flux would be totally dominated by the non-thermal,
featureless nuclear emission, which would make any redshift estimation a
very difficult task (see details in \cite{pad02} and Fig. 7 of
\cite{gio05}).  In other words, high-power HFSRQ sources would lack a
redshift (which is indeed the case for many BL Lacs with featureless
spectra) and therefore we would have no way of knowing that they are at
high power. 

The situation of the search for high-power--high-$\nu_{\rm peak}$ blazars,
the so-called HFSRQ, can then be thus summarised:

\begin{enumerate}

\item HFSRQ have been found and they make up $\sim 10\%$ of the FSRQ in
DXRBS, which is both X-ray- and radio-selected. The previously noted
absence of these sources was due to the fact that the majority of FSRQ
samples had been radio-selected and that no X-ray survey had looked for
FSRQ.

\item Despite the fact that these sources have $\nu_{\rm peak}$ values way
above those reached by LBL and well within the HBL range, their maximum
$\nu_{\rm peak}$ is $\sim 0.1$ keV, below the typical ($\sim 1$ keV), and
well below the largest ($\sim 100$ keV), values for HBL.

\item It is still not clear if the fact that HFSRQ do not reach the extreme
$\nu_{\rm peak}$ values of HBL is indicative of the fact that, after all,
there might be an intrinsic, physical limit to this parameter. An
alternative scenario is one where really high-power--high-$\nu_{\rm peak}$
blazars have their thermal emission swamped by the non-thermal, featureless
jet emission, which makes their redshift determination impossible. This
explanation for the the fact that presently known HFSRQ are of relatively
low power cannot be ruled out.

\end{enumerate}

In this respect, the discovery of a very powerful ($L_{\rm x} \sim
10^{47}$ erg/s) $z \sim 4$ FSRQ with a likely $\nu_{\rm peak} \ga 50$ keV
is extremely interesting and worth following up \cite{gio06b}. 

\begin{figure*}
\centering
  \includegraphics[width=0.75\textwidth]{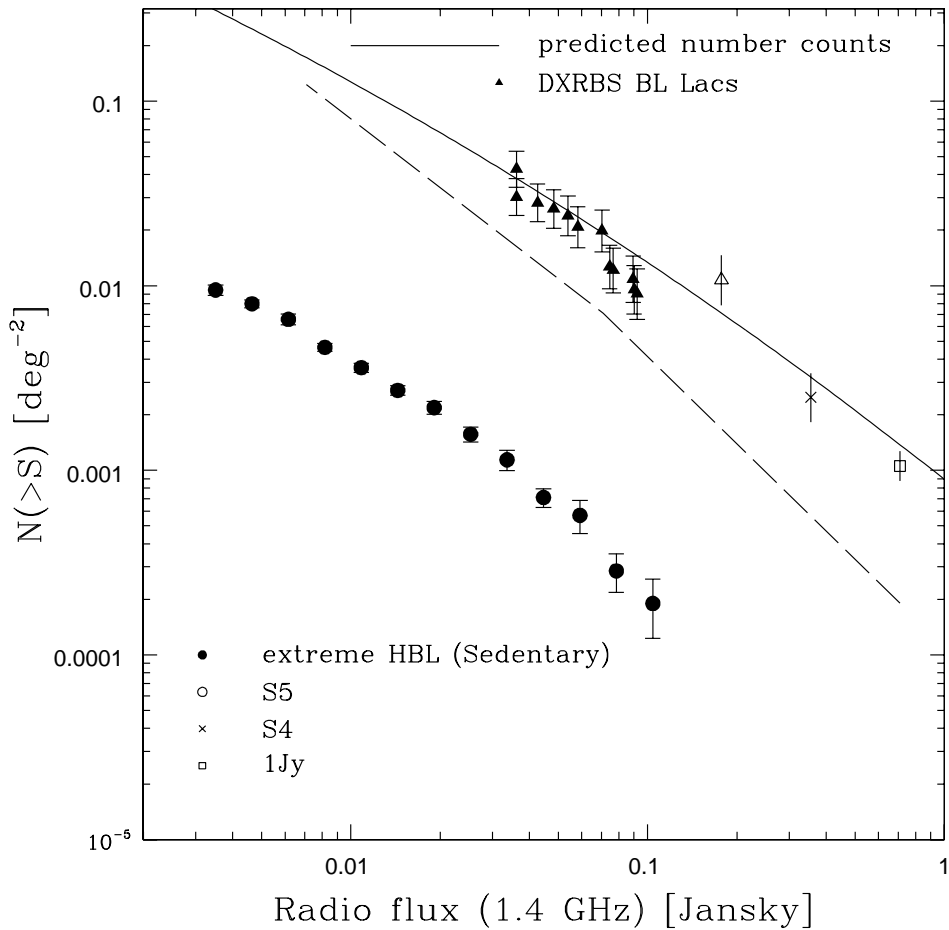}
\caption{The integral radio number counts for the extreme HBL Sedentary
sample at 1.4 GHz (filled circles). The solid line represents the expected
radio counts for all types of BL Lacs estimated from the radio luminosity
function of \cite{pg95}, while the dashed line shows the predictions of the
blazar sequence for all HBL \cite{fos97}. The BL Lac surface densities from the
1 Jy (open square), S4 (cross), and S5 (open triangle) are also shown,
together with the DXRBS number counts \cite{pad06}. All data apart from the
HBL have been converted from 5 GHz assuming $\alpha_{\rm r} =
-0.27$. Adapted and updated from \cite{gio99}.}
\label{sedent}       
\end{figure*}

\section{Counting Sources: are HBL more numerous than LBL?}

The last test to be carried out regards the relative fraction of the BL Lac
subclasses. Namely, are HBL more numerous than LBL, as predicted by the
blazar sequence?

The problem here is that the selection band affects the selected objects,
in the sense that X-ray selection finds mostly HBL, while radio selection
finds mostly LBL. Since the large majority of BL Lac samples are either
X-ray or radio selected (even optically selected ones, see \cite{col05},
have their biases, as the optical band would preferentially select HBL),
one would need an unbiased selection method, say a volume-limited
sample. Since we are still very far from having such an unbiased BL Lac
sample, one needs to make some assumptions and then predict the relative
fraction of HBL and LBL in X-ray- and radio-selected samples.

Let us assume that the blazar sequence is indeed a valid representation of
the truth and that, therefore, HBL are intrinsically the most numerous BL
Lac subclass. Therefore, although initially radio selection favours LBL,
the fraction of HBL in the radio band will have to increase at lower fluxes
until HBL become the majority at the faintest fluxes. In the X-ray band, on
the other hand, HBL are the dominant class and their fraction is expected
to be basically constant. Indeed, detailed predictions under a scenario
where $\nu_{\rm peak}$ has an an inverse dependence on bolometric power, as
required by the blazar sequence, have been worked out \cite{fos97} and
conform to these simple arguments.

Fig. \ref{sedent} shows the integral radio number counts for the HBL
Sedentary sample at 1.4 GHz. For lack of deeper samples, these were
compared by \cite{gio99} to the counts for all BL Lacs predicted by unified
schemes and based on the radio luminosity function fitted to the 1 Jy
sample assuming no evolution (Fig. 1 of \cite{pg95}). We can now
compare the HBL counts also with the observed BL Lac counts from DXRBS
\cite{pad06}. Fig. \ref{sedent} shows that the fraction of extreme HBL is
constant, as the HBL counts are parallel to the total ones. This is at
variance with the predictions of the blazar sequence, according to which
this fraction should increase at lower fluxes (dashed line in Fig.
\ref{sedent} \cite{fos97}).

The fraction of all (and not only extreme) HBL as a function of radio flux
could not be studied until the completion of our own DXRBS sample. We show
that the ratio LBL/HBL is basically constant and $\sim 6$ \cite{pad06},
while at the radio fluxes reached by DXRBS a value $\sim 2$ would be
expected \cite{fos97}. The HFSRQ fraction is also roughly constant with
radio flux. Again, this goes against what predicted by the blazar sequence.

\begin{figure*}
\centering
  \includegraphics[width=0.75\textwidth]{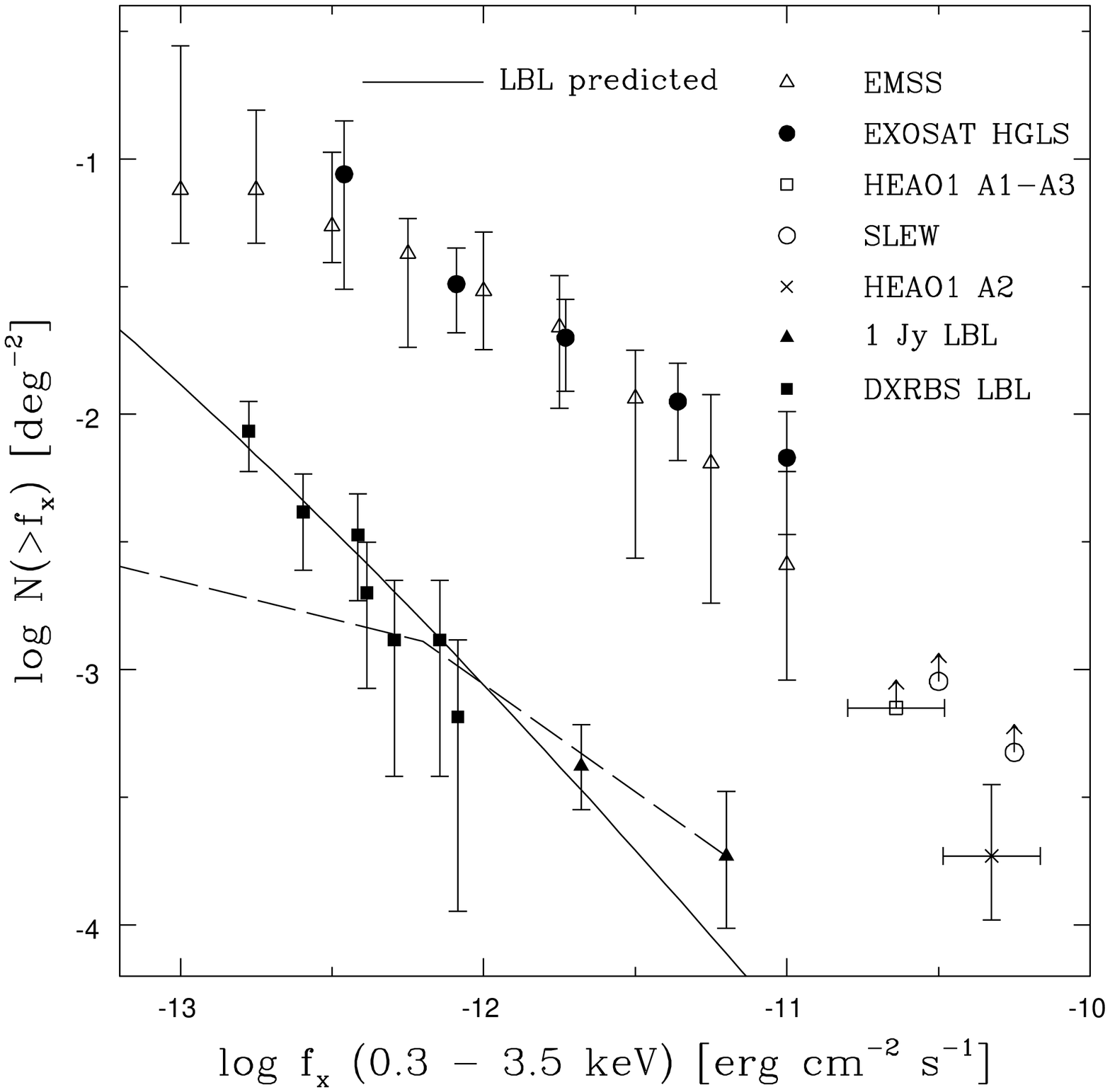}
\caption{The integral X-ray number counts for BL Lacs (\cite{pad06}, adapted from
\cite{pg95}). Data for five X-ray selected samples are shown. Filled
triangles represent the bivariate X-ray counts for the 1 Jy LBL with
$f_{\rm x} \ga 3 \times 10^{-12}$ erg cm$^{-2}$ s$^{-1}$, while filled
squares show the DXRBS LBL with $f_{\rm x} \ge 1.6 \times 10^{-13}$ erg
cm$^{-2}$ s$^{-1}$ \cite{pad06}. In both cases these define complete, X-ray flux limited
LBL samples. The solid line represents the X-ray number counts for LBL
predicted by \cite{gio94} and revised by \cite{pg95}, while the dashed line
shows the predictions of the blazar sequence \cite{fos97} normalized at
high fluxes.}
\label{xraycounts}       
\end{figure*}

We note that a completely independent argument against the strong increase
in the HBL fraction at low radio fluxes required by the blazar sequence can
be made \cite{gio06a}. In fact, were that the case, the blazar contribution
to the soft X-ray background, estimated around $\sim 12\%$ and mostly due
to the synchrotron component in HBL, would be much larger and inconsistent
with observational data.

The dependence of the fraction of LBL in an X-ray selected sample could
also not be studied to fluxes deep enough until DXRBS came into the
scene. This is shown in Fig. \ref{xraycounts}, which shows the integral
X-ray number counts for BL Lacs, adapted from \cite{pg95}, compared to our
best estimate of the integral number counts of LBL in the X-ray band \cite{pad06}. The
solid line represents the X-ray number counts for LBL predicted by
\cite{gio94} and revised by \cite{pg95} on the assumption that HBL
represent $\sim 10\%$ of the BL Lac population, while the dashed line shows
the predictions of \cite{fos97} normalized at high fluxes. The figure shows
that the fraction of LBL in the X-ray band increases at lower fluxes, as
expected if they were the dominant population. The blazar sequence, on the
other hand, predicts LBL to make up a constant fraction of the total, as
shown by the dashed line, in contrast with observations.

\section{Summary}

I have investigated the validity of the blazar sequence and tested its
predictions against recent observational data. My main conclusions are as
follow:

\begin{enumerate}

\item There is no anti-correlation between radio power and synchrotron peak
frequency in blazars, once selection effects are properly taken into
account. Furthermore, outliers to the originally proposed sequence have
been found, both in the low-power--low-$\nu_{\rm peak}$ and
high-power--high-$\nu_{\rm peak}$ regions.

\item The "missing" class of flat-spectrum radio quasars with synchrotron
peak frequency in the UV/X-ray band, whose existence is not expected within
the bla\-zar sequence, has been found.

\item Contrary to the predictions of the blazar sequence, all observational
data are consistent with the idea that the HBL subclass makes up a small
($\approx 10\%$) minority of BL Lacs.

\item Based on all of the above, the blazar sequence in its simplest form
cannot be valid.

\item The point remains, however, that the maximum synchrotron peak
frequency of FSRQ appears to be $\sim 10 - 100$ times smaller than that
reached by BL Lacs (but see \cite{gio06b} for a possible exception to this
rule). Is this telling us something about jet physics or is it still a
selection effect due to the fact that for really high-power--high-$\nu_{\rm
peak}$ blazars it might be hard to get a redshift estimate? This question
could be answered by the detection of high-power, moderately high-redshift
TeV blazars.

\end{enumerate}

\begin{acknowledgements}
Most of the work reported here has been done in collaboration with, amongst
others, Paolo Giom\-mi, Hermine Landt, Eric Perlman, Luigi Costamante, and 
Gabriele Ghisellini. Thanks to Paolo and Hermine for reading this paper and
providing useful comments. 
\end{acknowledgements}



\end{document}